\documentclass[conference]{IEEEtran}
\usepackage[utf8]{inputenc}
\usepackage[T1]{fontenc}
\usepackage{booktabs}

\usepackage{cite}

   \usepackage{graphicx}
\usepackage[cmex10]{amsmath}
\usepackage{amsfonts}
\usepackage[tight,footnotesize]{subfigure}
\usepackage{url}

\begin{document}
%
\title{Performance Evaluation\\of Biometric Template Update}

\author{\IEEEauthorblockN{Romain Giot and Christophe Rosenberger}
\IEEEauthorblockA{Université de Caen, UMR 6072 GREYC\\
ENSICAEN, UMR 6072 GREYC\\
CNRS, UMR 6072 GREYC\\
Email: romain.giot@ensicaen.fr\\
Email: christophe.rosenberger@ensicaen.fr}
\and
\IEEEauthorblockN{Bernadette Dorizzi}
\IEEEauthorblockA{Institut T\'el\'ecom; T\'el\'ecom SudParis\\
UMR 5157 SAMOVAR\\
Email: bernadette.dorizzi@it-sudparis.eu}}


%


\maketitle


\begin{abstract}
  Template update allows to modify the biometric
reference of a user while he uses the biometric system. 
With
such kind of mechanism we expect the biometric system uses
always an up to date representation of the user, by capturing
his intra-class (temporary or permanent) variability. 
Although
several studies exist in the literature, there is no commonly
adopted evaluation scheme. This does not ease the comparison of
the different systems of the literature. In this paper, we show that
using different evaluation procedures can lead in different, and
contradictory, interpretations of the results. We use a keystroke
dynamics (which is a modality suffering of template ageing quickly) template update system on a dataset consisting of
height different sessions to illustrate this point. Even if we do
not answer to this problematic, it shows that it is necessary to
normalize the template update evaluation procedures.

\end{abstract}

%
\IEEEpeerreviewmaketitle

\section{Introduction}

Template update is an active research field whose aim is to
update the biometric reference of individuals while using the
biometric system.
Even if the reason of using template update systems are various
(template ageings, noisy acquisitions, lack of samples during
enrollment, ...), the expected result is always the same: the improvement of
the recognition performance.

Template update mechanisms may vary depending of different factors (which
are not directly subject of this work, as we are interested on
the evaluation of this mechanism):
\begin{itemize}
    \item The choice of the of update criteria (threshold, graph
    based, ...).
    \item The periodicity of the template update (online and
    batch, or offline, at various frequencies).
    \item The working mode of the template update system
    (supervised or semi-supervised): in the first case, we guaranty
    no impostor data has been used for the template update.
    \item The template update mechanism (mainly the employed
    method used to modify the biometric references).
\end{itemize}

A very nice work~\footnote{although specific to keystroke
dynamics}, exposes the various points of differences to specify in
the studies~\cite{seeger2011how} (they argue that these
informations are mainly
missing in studies).
Nevertheless, this work does
not explore the performance evaluation procedure computation
(they give information about the way of evaluating the system,
but not on the way of computing the error rate). It is necessary
to quantify the performance evolution using such kind of
mechanism. We will show that different performance computing methods lead to different interpretations of the results. In
this work, we present the differences in the various template
update (or related) evaluation schemes in the literature. We
do not emphasize on the template update mechanisms. We
raise the questions that must be answered by the template
update community in order to allow an easy evaluation and
comparison of the template update mechanisms.

The paper is organised as following.
Section~\ref{sec_Available Databases} briefly presents the datasets used
in the literature for works on template update.
Section~\ref{sec_Existing evaluation schemes} presents the
different ways encountered in the literature to evaluate the
template updating schemes.
Section~\ref{sec_illustration} illustrates the problem of not
having a common evaluation methodology in the template update
studies.
Section~\ref{sec_Remaining Questions} raises various open
questions on template update evaluation methodology.

\section{Available Public Databases} 
\label{sec_Available Databases}

Studies on template update require adequate
datasets. Various datasets have been used in the literature.
They all differ in number of subjects, number of samples
per subjects, number of sessions, time difference between
the youngest and oldest sample, type of variability\ldots
The following datasets have been used in the literature in
template update works or in studies analysing the variability of
samples through time:

\begin{itemize}
    \item 2D face recognition: there are several datasets for face
    recognition. In this case, the variabilities are mainly due to
    pose or illumination differences, but few datasets allow the
    study of templates ageing by
    capturing data on a very long period while having a lot of
    users.
    \begin{itemize}
        \item The \emph{Equinox Face Dataset}~\cite{equinox} is often
        used but does not seem to be yet freely available. The
        number of individuals and samples varies between
        studies (they do not use the same subset).

        \item The dataset \emph{MORPH}~\cite{ricanek2006morph} has been
        used in several studies. Once again, the number of
        individuals and samples varies in studies.

        \item The \emph{UMIST Face database} contains 564 images
        of 20 individuals. Most studies in the state of the art do not use the whole
        set.

        \item The \emph{AR}~\cite{martinez1998ar} contains several
        color images of 120 individuals captured on two
        sessions.

        \item Drygajlo \emph{et al.}~\cite{drygajlo2009q} used youtube's videos of
        people providing their face each day during three years in
	general.
        The timespan is superior to the other
        datasets, but the number of users is very low and no ground
        truth is available (automatic image extraction can be
	erroneous, nothing proves that pictures are presented in
	chronological order,\ldots).

	\item \emph{VADANA}~\cite{gowri2011vadana} is the most recent dataset designed
	especially for template update in face recognition
	systems. 43  subjects have in average 53 pictures, delta
	between two pictures of an individual can be of several
	years. This dataset has more intra-class comparison than
	other long term datasets.

    \end{itemize}

    \item 3D face recognition. 
        The \emph{Face Recognition Grand Challenge (FRGC)
        Experiment 3}~\cite{phillips2005overview} provides 3D
        faces linked to color information.
        Dataset is splitted in a training set of 270 individuals and
        a testing in of 410 individuals.

    \item Fingerprint recognition.
        The dataset~\cite{fvc2002} comes from the
        competition  ``Fingerprint Verification
        Competition''.
        Four different sub-datasets are available. 
        Each of them contains 110 fingers with 8 samples per
        finger. 
	This dataset is not appropriate to study variation through
	time, but it is interesting because of the high
	intra-class variability of
	users~\cite{freni2008replacement}.

    \item Keystroke Dynamics.
    \begin{itemize}
        \item The \emph{GREYC keystroke}~\cite{giot2009benchmark}
        dataset has been captured among 5 distinct sessions
        with 100 individuals.

        \item The \emph{DSN2009}~\cite{killourhy2009cad} has been
        captured amoung 8 distinct sessions with 51 individuals.
    \end{itemize}

    \item Handwritten signature.
        The dataset \emph{MCYT-100}~\cite{ortega2004mcyt} is
        a multimodal biometric database (fingerprint and handwritten
        signature) which has been used to verify the reliably of
        extracted features through
        time~\cite{houmani2009assessing}.
\end{itemize}

We can see there are various datasets available for several different biometric
modalities ; they are summarised in the table~\ref{tab_datasets}.
Most dataset are related to 2D face recognition which is a
morphological modality which hold less variability than any
behavioral biometric.
The properties of these datasets are really different.
Few of them have been captured in a long timespan.
They are more useful to analyze the intra-class variability
due to temporary variations than template ageing.

In the next section, we present the existing evaluation schemes for template update algorithms.

\begin{table}[!tb]
  \centering
  \caption{\label{tab_datasets}Summary of the datasets used in the
  literature. Figures are
related to studies using the dataset and may be different from
the real value of the dataset. We can see than few of them
seem appropriate for template update studies.}
\begin{tabular}{@{}llll@{}}
\toprule
Database       & \# users & \# samples  & \# sessions \\
\midrule

\addlinespace
\multicolumn{3}{l}{\textbf{2D face}}\\
EQUINOX        & 40-50    & 20-100      & -\\
MORPH          & 14       & > 20        & -\\
UMIST          & 20       & 25-55       & -\\
AR             & 120      & 26          & 2\\
YOUTUBE videos & 4        & 1200        & 1200\\
VADANA         & 43       & $\approx$53 & -\\
\addlinespace
\multicolumn{3}{l}{\textbf{3D face}}\\
FRGC-EXP3      & 410+270  & 1-22        & -\\
\addlinespace
\multicolumn{3}{l}{\textbf{Fingerprint}}\\
FVC2002        & 110      & 8           & 1\\
\addlinespace
\multicolumn{3}{l}{\textbf{Keystroke dynamics}}\\
GREYC2009      & 100      & 60          & 5\\
DSN2009        & 51       & 400         & 8\\
\addlinespace
\multicolumn{3}{l}{\textbf{Handwritten signature}}\\
 MCYT-100      & 100      & 25          &5 \\
\bottomrule
  \end{tabular}
\end{table}

\section{Existing Evaluation Schemes} 
\label{sec_Existing evaluation schemes}

Few template update studies exist in the literature.
In this section, we present the different evaluation protocols
found in the literature, using
datasets separated in several sessions (also called \emph{batch}
in some studies), or not.
We also present the different ways of presenting the
queries to the biometric references.

\begin{figure*}[!tb]

  \includegraphics[width=\linewidth]{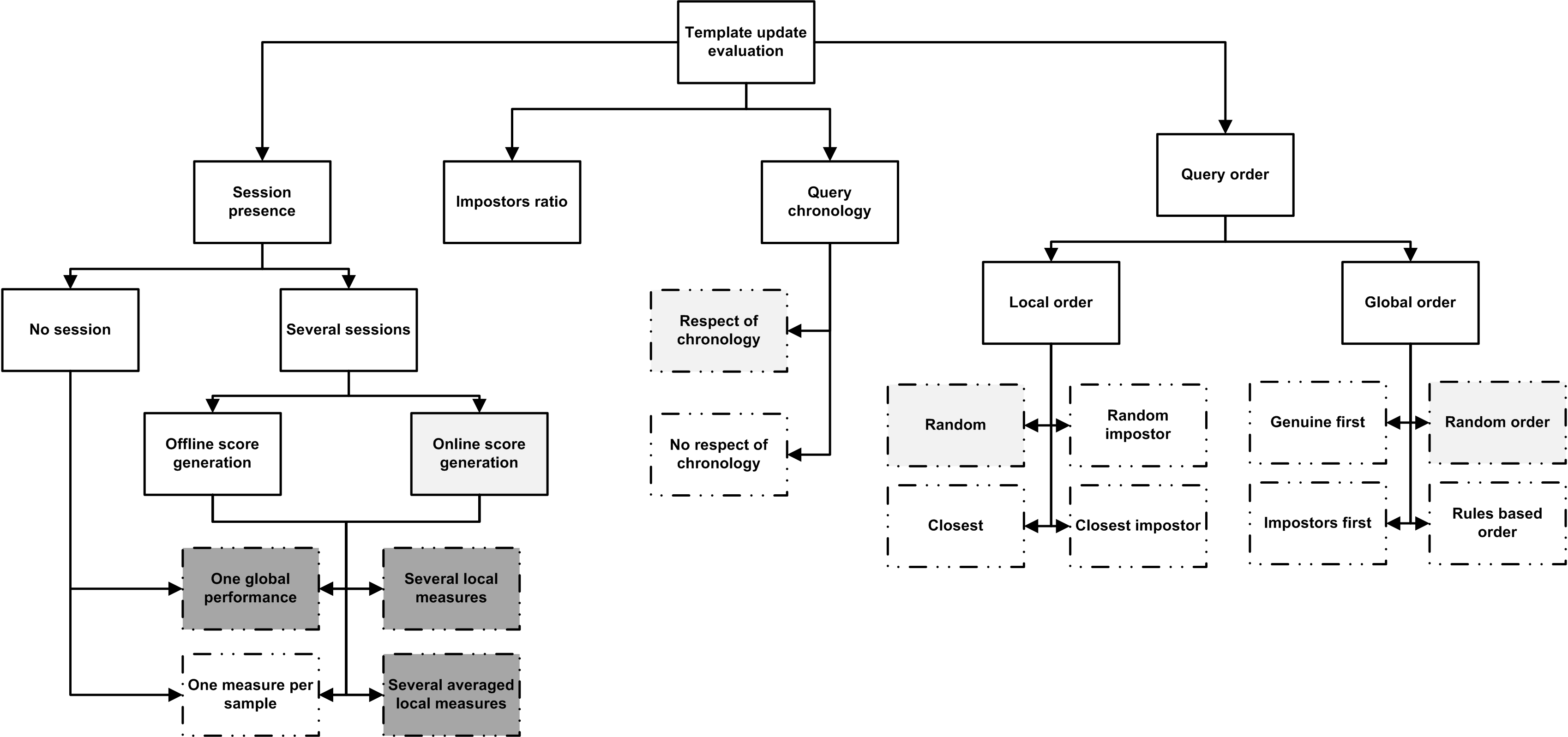}
  \caption{Summary of all the possible variabilities in a template
  update evaluation. 
  Dotted nodes represent the possible configuration values, while
  nodes with a straight line represent the configuration types.
  Dark gray nodes represent the variant factors in
  Section~\ref{sec_illustration}, while light gray nodes represent the fixed factors in Section~\ref{sec_illustration}. 
  \label{fig_summary}}

\end{figure*}

\subsection{Studies With Several Sessions} 
\label{sec_Studies per sessions}

Using dataset providing several capture sessions allows computing error rates
specific to sessions.
This way, we can track the evolution of the template update through time.
Curiously, it is only recently that this kind of evaluation has
been encountered~\cite{giot2011analysis, rattani2011self}.
Maybe, such kind of studies is not common because the data acquisition is not
very straightforward and too much time consuming.

In such kind of studies, the first session is used to compute the biometric
reference of each user, while the next ones are used to apply the template update
mechanism and evaluate the update procedure.
We can observe two main evaluation processes:
\begin{itemize}
  \item An \emph{online} order where the comparison score of
  the query against the reference is used to compute the evaluation
  measure (and is not only used in the template update mechanism).
  \item An \emph{offline} order where the comparison score of
  the query against the reference is not used to compute the
  evaluation measure. When the whole query set of the session is
  consumed, the entire query
  set of the next session is used to evaluate the new biometric
  references. Following this step,
  this set is then used for the template update
  procedure.
\end{itemize}

Our personal investigations suspect that these two evaluation
schemes do not give
fundamentally different results, and that the online scheme must be favored to
the offline one because: 
\begin{enumerate}
  \item  it simplifies the evaluation
  procedure, 
  \item it avoids unnecessary computations,
  \item it
  produces an additional session result (as the latest session does
  not need an additional session to be evaluated).
\end{enumerate}
We have also met two different ways of presenting the results:
\begin{itemize}
    \item One performance measure per
    session~\cite{giot2011analysis} computed with one of the
    previously presented methods. 
    This gives result specific to each sample of the session.
    \item One performance measure per session computed by averaging the
    performance of the current session and the previous
    ones~\cite{rattani2011self}.
    Authors argue this is important because the error rate depends
    too much on the used test.
    This smoothing reduces the error rates in comparison to the previous
    method.
    \item One global performance computed with the whole set of
    scores~\cite{seeger2011how}.
      \end{itemize}

\subsection{Studies Without Any Session} 
\label{sec_Studies without sessions}

Most template update studies use datasets with no session, but samples
captured in a more or less long period.
We can observe two main evaluation procedures:
\begin{itemize}
    \item Separation of the dataset in two (or three)
    sub-datasets, which act
    as if they were two sessions dependant datasets. 
    In this case, the applied procedures are similar to the previously
    presented ones~\cite{roli2008abs}.
    In this case, we have only one performance measure for the template
    update system on the entire dataset.
    \item Computation of the biometric performance at any time, by modeling
    its behavior~\cite{poh2007method}.
    Note that this method has been illustrated in order to observe the
    behaviour of a biometric system using no template update system.
    But, we think it can be used in order to evaluate the
    performance of an online template update system.
\end{itemize}


\subsection{Query Presentation Order} 
\label{sub_Query Presentation Order}
Another factor, in the template studies, is the query samples
presentation order.
We think this information belongs to the evaluation procedure and
not directly the template update system, because performance is 
dependent of them.

In~\cite{seeger2011how}, authors make the distinction between
global and local orders.

\subsubsection{Global order} 

The differences can be:
\begin{itemize}
    \item The proportion of impostor samples:
    this is a very important information, as this factor highly impacts the
    performance: many impostor samples increases the probability of including
    impostor samples in the biometric references and decreases the performance.
    This information may be unavailable, fixed at one specific value
    (50\% for example), or several ratios can be
    specified~\cite{giot2011analysis}.
    
    \item The presentation order of the different types (genuine
    or impostor) of samples. 
    This is also an important information, as this factor can also impact the
    performance by driving the probability of doing wrong template updates.
    We mainly meet three different behaviors.
    Depending on the studies,
    one~\cite{giot2011analysis,rattani2011self} or all~\cite{ryu2006template} of them can
    be present.
    The behaviors are:
    \begin{itemize}
      \item Presenting the \emph{genuine samples first}.
        All the genuine samples are presented before the impostor samples.
        Before presenting the first impostor query, the biometric reference might
        already be highly specialised to efficiently recognize genuine queries
        and reject impostor queries.
        We expect really good recognition rates and few impostor samples
        inclusion in the biometric reference.

	\item Presenting the \emph{impostor samples first}.
        All the impostor samples are presented before the genuine samples.
        Before presenting the first genuine query, the biometric reference migh
        already be highly un-specialised and performs poor results
	(by having
        included too many impostor samples and no genuine ones to counterbalance
        that).
        We expect quite poor recognition rates and a lot of impostor samples
        inclusion in the biometric reference.

	\item \emph{Random order presentation}.
        No specific order is preferred.
        The presentation order is totally random (although controlled by the
        impostor ratio).
        A good template update system should include a lot of genuine samples and few
        impostor samples, while a bad template system includes a lot of impostor
        samples and few genuine samples.
        Performances are averaged but probably more realistic than
	in the first two cases.
	Of course, this must be done for different impostor ratios.

	\item \emph{Rules based order}. The order is directed by a set
of rules to follow. Such kind of order is problem
specific.

    \end{itemize}
\end{itemize}

\subsubsection{Local Order} 
\label{ssub_Local Order}
The local order pays attention to the order
of presenting impostors samples.
\begin{itemize}
  \item \emph{Totally random}. A random sample from a random impostor is
  selected.
  \item \emph{Closest}. The closest sample (among all the samples of all
  the impostors) from the biometric reference is chosen.
  \item \emph{Random impostor}. An impostor is chosen randomly. His
  samples are used, in a chronological order for behavioral
  biometrics, until another impostor is selected.
  \item \emph{Closest impostor}. The impostor closer to the biometric
  reference is selected. His samples are used, in a chronological order for behavioral biometrics, until another
  impostor is selected.

\end{itemize}


\subsection{Query Chronology} 
\label{ssub_Query chronology}
The last important information, regarding the evaluation, is the
respect, or not, to the chronology information.
When this information is presented, we met two kinds of papers:
\begin{itemize}
  \item \emph{No chronology respect}.
    In these papers, samples chronology is not respected. 
    It means that a query $B$ tested against a biometric reference after a query
    $A$ can be younger than $A$.
    In average:
    \begin{equation}
    \mathbb{P}(age(A)<age(B))=\mathbb{P}(age(B)<age(A))
    \end{equation}
    \noindent with $\mathbb{P}(e)$ the probability of the event
    $e$ and $age(s)$ the age of the sample $s$.
    This procedure is the most common in the literature whereas it can only be
    efficient if we assume that the template variability is not related to
    ageing but other factors.
    This is of course false for the behavioral modalities and not always true
    for the morphological ones.

    \item \emph{Respect of the chronology}.
    The assumption is that biometric sample variability is also
    related to ageing of the biometric data (whatever the reason).
    Genuine samples are always presented by chronological order, but not
    necessary impostor samples:
    \begin{eqnarray}
    \mathbb{P}(age(A)<age(B))=1 \\
    \mathbb{P}(age(A)\geq age(B))=0
    \end{eqnarray}
\end{itemize}


From this review of the literature, we observe that all studies
use different protocols, and, that up to now, no standard
evaluation procedure exits.
Figure~\ref{fig_summary} summarised the various points subject of
variations.
It could not be a problem if all
these points are indicated in studies~\cite{seeger2011how},
because they can be representative of different but useful scenarios. However,
when the performance evaluation procedure differs, it can hold
to no similar results.

We will illustrate the problem that
such a situation can provide in Section~\ref{sec_illustration}.

\section{Illustration} 
\label{sec_illustration}
\begin{figure*}[!tb]
  \centering
  \subfigure[Template update system 1]{\includegraphics[width=.45\linewidth]{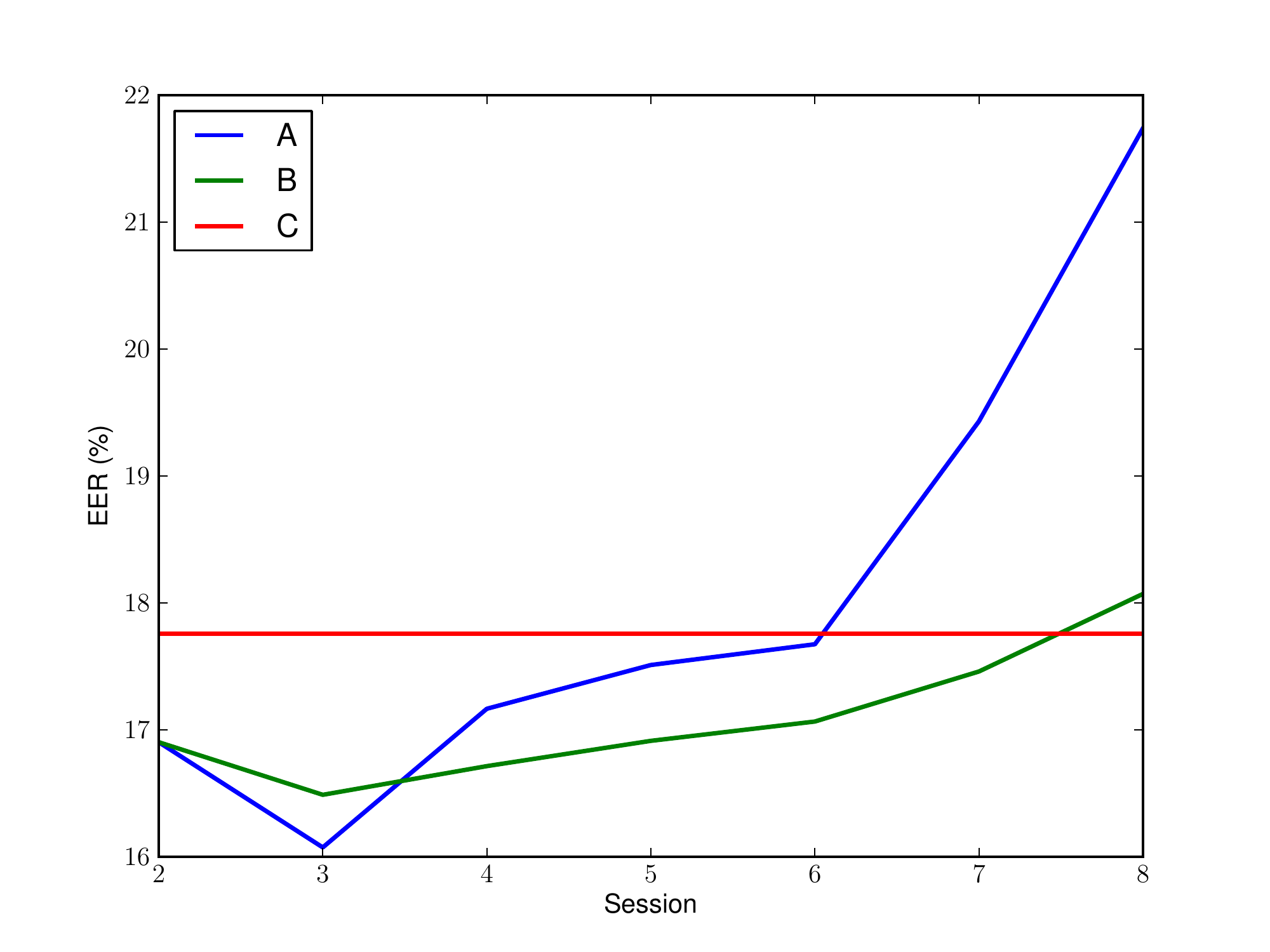}}
  \subfigure[Template update system 2]{\includegraphics[width=.45\linewidth]{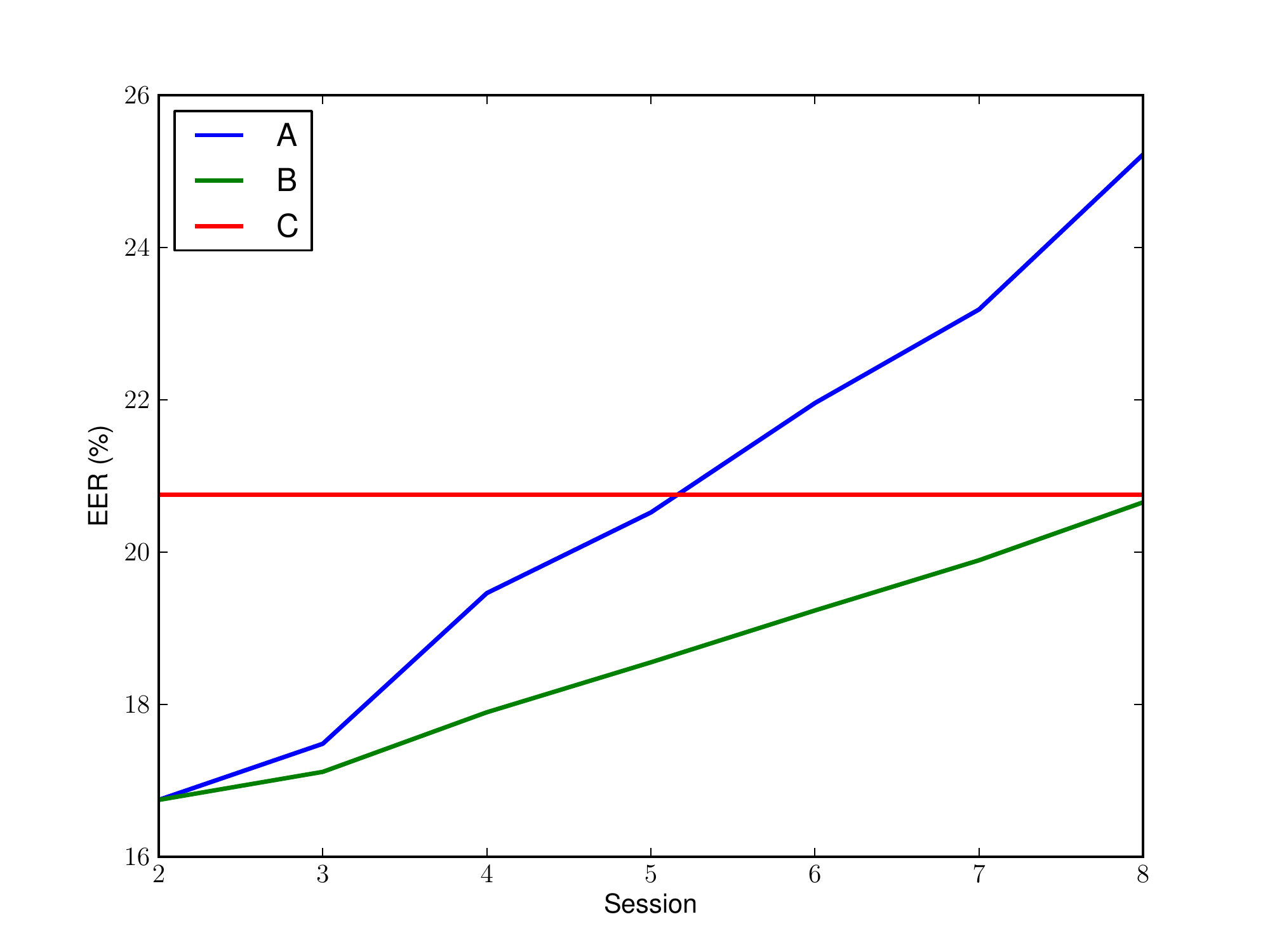}}
  \caption{\label{fig_perf_eer}
  Performances depending on the evaluation method on the same
  score set. (for the definition of A, B, C see
  Section~\ref{sec_illustration}).}
\end{figure*}

The previous section presents the various differences in
the evaluation procedure of a template update system. The
variation of one factor holds to another testing scenario. We
have not discussed about the evaluation of these scenarios.

In this example, we are interested in the evaluation of a
template update mechanism~\cite{giot2011analysis}  for a keystroke
dynamics~\cite{giot2011keystroke} system using the
Equal Error Rate (EER) as the evaluation metric.
We are not interested in the characteristics of the template update
system.
This system, which is presented
in~\cite{giot2011analysis} aims at applying a semi-supervised
update based on an update threshold.
We have selected two different configurations of the template
update system:
\begin{itemize}
  \item  \emph{System 1}: a scenario where the update threshold (distances
  can be negative) is $-0.2$.
  \item \emph{System 2}: a scenario where the update threshold is
  $-0.3$.
\end{itemize}

The following fixed parameters are used for the evaluation:
\begin{itemize}
  \item The dataset~\cite{killourhy2009cad} provides \emph{8 sessions}. The ways of
  computing the performance measure are presented later.
  The first session serves to compute the initial biometric
  reference.
  The other sessions serve to update the reference and compute the
  performance of the updating system.
  \item We compute the scores for each session in an \emph{online}
  way.
  \item The \emph{impostor ratio} is $30\%$.
  \item As it is a behavioral modality, we \emph{respect the
  chronology}.
  \item The global order of presentation of genuine or impostor
  samples is \emph{random}.
  \item The local order of presentation of impostor samples is
  random too.
\end{itemize}

This configuration allows us to compute the comparison scores
while the system is updating.
In addition of these fixed parameters, we have chosen to select
three different ways of computing the performance value from these
comparison scores.
Three different evaluation procedures are applied (the selected performance
indice is the EER):
\begin{itemize}
  \item \emph{Performance evaluation A}. As done in our previous
  work~\cite{giot2011analysis} where the scores of the current session are
  used to compute its performance.
  
      \begin{equation}
	      A_i = \makebox{EER}\left(scores_i\right)
	      , \quad \forall i , 2 \leq i \leq S
	    \end{equation}

	    \noindent with $S$ the number of sessions,
	    $\makebox{EER}(\cdot)$ the EER computing function and
	    $scores_i$ the scores computed at session $i$ (intra
	    and inter comparisons).
	    We have one EER per session.

	    \begin{equation}
	      \mathbf{A} = [A_2, \ldots, A_{S}]
	    \end{equation}

  \item \emph{Performance evaluation B}. As done in~\cite{rattani2011self} where performance of current
   session is
   computed by the mean of all the previous session
   performance (including the current one).

       \begin{equation}
	      B_i = \frac{1}{i-1} \sum_{j=2}^i
	      \makebox{EER}\left(scores_j\right)
	      , \quad \forall i , 2 \leq i \leq S
	    \end{equation}

	    \noindent We also have one EER per session.

	    \begin{equation}
	      \mathbf{B} = [B_2, \ldots, B_{S}]
	    \end{equation}

   \item \emph{Performance evaluation C}. As done in~\cite{seeger2011how}
   where only one measure is computed. In the present case, we merge
   all the scores of all the sessions in one global set and compute the
   performance measure on this set.

   	    \begin{equation}
	      C = \makebox{EER} \left( \bigcup_{i=2}^{S} scores_i \right)
	    \end{equation}

	    \noindent We have one EER for the whole interval.
	    To compare it easily with the two other methods we
	    duplicate it the number of test sessions times.

	    \begin{equation}
	      \mathbf{C} = [\underbrace{C, \ldots,
	      C}_{S-1\makebox{ values}}]
	    \end{equation}

\end{itemize}
This evaluation procedure is repeated ten times and the results are averaged (as the
process is stochastic due to the impostor choices and order).

Figure~\ref{fig_summary} presents in light gray this fixed
configurations and in dark gray the varying configurations.
Figure~\ref{fig_perf_eer} presents the performance, on exactly the same
set of scores, of the three evaluation schemes A, B and C.
Although globally, the three different evaluations show that system
1 is better than system 2 (better update involving lower EER), we
can propose totally different interpretations of the updating
system, depending on the chosen evaluation scheme:
\begin{itemize}
  \item \emph{Performance evaluation A}.
	Performance of system A decreases fast with time, the
	template update system does not perform well.
	The template update system must be improved, or the
	biometric modality has a very low permanence.
  \item \emph{Performance evaluation B}.
	Performance of system B decreases with time, but the amount of
	decreases is not really important, the template update
	system is not too bad.
	The template update is not perfect (there is a performance
	decrease) but it takes quite well the ageing into account.
  \item \emph{Performance evaluation C}.
	Performance of system C is averaged, but we cannot know if
	it is because of template ageing, because of a bad
	algorithm or because of a bad dataset.
\end{itemize}

As no performance measure of a system without template update is
presented, we cannot compare the template update systems against
the baseline classifier.
By the way, the performance evaluation of a system without
template update would hold the same performance evaluation
problem.
The performance evaluation C brings less information than the two
other ones.
So it must be avoided, because we lack the temporal information
which is the most important one.
However performance evaluation A and performance evaluation B
track temporal evaluation, but give different interpretations.
Which one is the most interresting or accurate?
In the next section, we raise the questions it would be
interesting to answer in order to normalize template update
evaluation.


\section{Open Questions} 
\label{sec_Remaining Questions}

All along this paper, we have analyzed the differences in the
evaluation protocols, one can encounter in the various 
biometric template update studies.
The variability found in all the protocols raise many open
questions:
\begin{itemize}
	\item What are the characteristics of an interesting dataset
    for such kind of studies?
    We have seen that there are several datasets available for the
    different modalities; they are
    different in their sample distribution.
    Few of them seem really interesting to be used in template
    update scenarios.
    It is important to know what are the interesting
    characteristics to respect in order to create new useful
    datasets.

    \item What is the best evaluation procedure in order to easily
    compare the systems without doing each time all the previous
    experiments from scratch ?
    The update evaluation procedure is not yet standardized and
    procedures are really different between studies.
    Maybe, it is interesting to create new metrics specific for
    such kind of problem.
    Some studies present the ratio of impostors included in the
    updated biometric reference, but other metrics could be
    interesting too.

    \item Is it more informative to work with datasets separated
    in several sessions, or with datasets captured in a longer
    period without more information ?
    We can suspect that:
    \begin{itemize}
    	\item In the first case, we have datasets with a small
	intra-class variability within sessions and a bigger
	variability between sessions.
	\item In the second case, we have datasets with in an
	intra-class variability homogeneously spread other time.
    \end{itemize}

\end{itemize}

Without answering these questions, it will be hard to homogenize
and compare the different studies on template update mechanisms.



\section{Conclusion}
We have presented the different template update evaluation schemes encountered
in the literature.
We can observe that there exist lots of different and incompatible ways to do it.
This hardly allows the comparison of template update mechanisms and their
understanding.
This asserts the request for the researchers of being very accurate while explaining the experimental
protocol in order to ease the reproducibility of the experiment.


\IEEEtriggeratref{2}


\bibliographystyle{IEEEtran}
%

\bibliography{template_update_evaluation}

\end{document}